\documentclass[12pt,a4paper]{iopart}
\usepackage{geometry}                
\usepackage[dvipdfmx]{graphicx}
 \usepackage{bmpsize}
\usepackage{amssymb}
\usepackage{epstopdf}
\usepackage[linewidth=.5pt]{mdframed}
\usepackage{color}
\usepackage{caption}
\usepackage{array,booktabs}
\newcommand{\cfig}[1]{Fig[{\ref{#1}}]}

\def\Black{}
\def\Red{}
\newcommand{\sig}{\mbox{\boldmath $\sigma$}}
\def\signed #1{{\leavevmode\unskip\nobreak\hfil\penalty50\hskip2em
  \hbox{}\nobreak\hfil(#1)%
  \parfillskip=0pt \finalhyphendemerits=0 \endgraf}}
\newsavebox\mybox
\newenvironment{aquote}[1]
  {\savebox\mybox{#1}\begin{quote}}
  {\signed{\usebox\mybox}\end{quote}}
\newcounter{quotecount}

\begin{document}
\title{Statistical Mechanics and the Ghosts of Departed Quantities}
\author{G. N. Ord}
\address{Department of Mathematics,
Ryerson University, Toronto, Ont. Canada.
}

\begin{abstract}
  Quantum mechanics appears to contain ghosts from both classical statistical mechanics and special relativity. On one hand, both the Dirac and Schr\"{o}dinger equations have classical analogs that emerge directly from classical statistical mechanics, unimpeded by major problems of interpretation. On the other hand, the formal analytic continuation that takes these classical equations to the quantum version introduces a velocity dependent phase. However, among classical theories, only in relativistic mechanics does one find path-dependent phase in the form of relativistic time dilation.   This paper explores the idea that if we start with statistical mechanics and special relativity we can discover  a version of the quantum algorithm and show that at least some of the resulting ghosts are direct descendants of those connected with the birth of the differential calculus.  
\end{abstract}

\section{Introduction}
Stuart Whittington was my Ph.D. supervisor in the  early 1980's. As all Stu's colleagues and former students know, SGW is unique. He has a singular talent for extracting clear and often beautiful mathematical models from complicated physical processes. He is an ideal mentor, providing insight and encouragement for successes and humour and perspective for failures. My time as his student was one  of  discovery and exploration,  inspired by Stu's intellectual curiosity that ranged over the full spectrum of scientific endeavour.

As Stu's  student, I looked forward to our periodic meetings. These evolved into
 a regular pattern. I would present the latest (often infinitessimal) progress on thesis work. Stu would gently but firmly extract whatever substance could be saved, while reminding me that only a finite amount of time was available for the completion of a thesis.  With this mandatory concession to the practicalities of  thesis work over, we would step back and  discuss whatever  questions occurred to us at the time. A common feature of these meetings was that they usually visited questions {\em about} science or mathematics rather than just staying within  particular mathematical models. 

For example, long before  statistical mechanics arose to mediate between opposing views  on the continuity of matter, there were objections to the calculus itself. Talking about the then relatively new differential calculus,  Bishop Berkeley had this to say about `fluxions':

\begin{aquote}{From The Analyst. \cite{Berkeley}  1734}{  $ \cdots$  they are neither finite quantities nor quantities infinitely small, nor yet nothing. {\em May we not call them the ghosts of departed quantities?}}
\end{aquote}

The motivation for the irony here was that the Bishop thought, rightly so as it turned out, 
 that scientists 
   had deeply held beliefs about the calculus that were not supported by the mathematical rigour of the time\footnote{The technical issue was about the use of infinitesimals. These were used directly in early formulations of the calculus but were subsequently discarded in favour of the more precise use of limits. It was not until the 1960's that  infinitesimal formulations were put on a more rigorous basis\cite{infiprimer}. }. He apparently felt that those who would mock the clergy for unsupported beliefs should be more circumspect in their own field.
 
 While real analysis  responded to Berkeley's criticisms concerning a lack of   mathematical rigour, modern statistical mechanicians arguably have a more practical appreciation of the spirit of Berkeley's ghosts than their colleagues in more esoteric areas of theoretical physics. Limits in statistical mechanics are often explicit and are usually understood to be a form of controlled approximation. The subject of this paper is an extension of conversations SGW and I had about this. 

As an historic example of a statistical mechanical perspective, consider the diffusion equation

\begin{equation}
\frac{\partial u}{\partial t} = D \frac{\partial^{2}u}{\partial x^{2}}.\label{eqn:diffn}
\end{equation}

\begin{figure}[t] 
   \centering
   \includegraphics[width=2in]{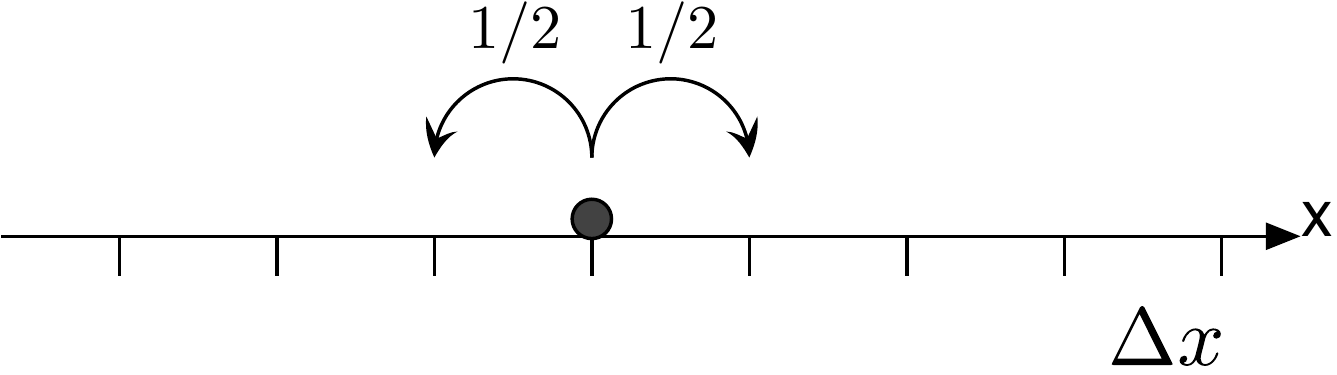} 
   \caption{A simple symmetric random walk on a lattice.\cite{CC0}}
   \label{fig:randomwalk}
\end{figure}

\noindent It is unlikely that any statistical mechanic would consider the equation  `fundamental' in any physical sense. Thinking of a random walk model of diffusion on a lattice, \cfig{fig:randomwalk}, one can quickly sketch an argument that indicates the origin of the equation {\em and its limitations.}
\noindent  If $u(x,t)$  represents a probability mass function for a particle executing a symmetric random walk on a square lattice with spacings $\Delta x$ and $\Delta t,$ then assuming that particles can only hop left or right with equal probability
\begin{equation}
u(x,t+\Delta t) =\frac{1}{2}(u(x-\Delta x,t)+u(x+\Delta x,t)) \label{eqn:mass}
\end{equation}
establishes a difference equation. 
This simply states that a particle is equally likely to have come from either prior lattice point; it represents conservation of mass. We might rewrite it as
\noindent 
\begin{equation}
u(x,t+\Delta t) -u(x,t)=\frac{1}{2}(u(x-\Delta x,t)+u(x+\Delta x,t))-u(x,t).
\end{equation}

\noindent  Now assuming that we can refine the lattice and that $u(x,t)$ is ultimately analytic,  the above form suggests a Taylor expansion. The zero-order terms on either side cancel leaving:

\begin{equation}
\frac{\partial u(x,t)}{\partial t} \Delta t =\frac{1}{2}\frac{\partial^{2} u(x,t)}{\partial x^{2}} (\Delta x)^{2}.
\end{equation}

\noindent  where we have dropped higher order terms. Rearranging this we get

\begin{equation}
\frac{\partial u(x,t)}{\partial t} =\left(\frac{(\Delta x)^{2}}{2 \Delta t}\right) \ \frac{\partial^{2} u(x,t)}{\partial x^{2}}. \label{eqn:d2}
\end{equation}
\noindent Here we see the emergence of the differential equation (\ref{eqn:diffn}) provided that the lattice may be refined in such a way that 
\begin{equation}
\left(\frac{(\Delta x)^{2}}{2 \Delta t}\right) \rightarrow D \label{eqn:D}
\end{equation}
\noindent throughout the refinement process. That this is the appropriate scaling for symmetric random walks follows from the fact that the mean square end-to-end distance of an $n$-step random walk is $O(n)$. 

From a physical standpoint  the partial differential equation emerges as a mathematical model for statistical reasons;  the statistical approach points to both the partial differential equation {\em and to its  limitations involving characteristic lengths and times.}  The justification for (\ref{eqn:D}) is the existence of  counting arguments about stochastic processes.  The mathematical limit this entails is sensible, it mimics what one might expect to find as a result of molecular motions, it can be directly checked and simulated. However we also know that it is only physically reasonable over a restricted range of scale. Specifically we would expect (\ref{eqn:D}) to mimic physical systems provided $\Delta x$ and $\Delta t$ stay above the characteristic scales of the mean free path and mean free time respectively. Below these scales we do not expect the diffusion equation to be a good model  of the results of physical Brownian motion. 
This discrepancy between the arguments that take us from (\ref{eqn:mass}) through to (\ref{eqn:D}) and the diffusion equation itself, illustrates part of the spirit of Berkeley's criticism. Here the partial differential equation and its solutions correspond to Berkeley's ghosts and the `departed' quantities contain the statistical reasoning that determines the context in which the PDE is a good mathematical model of physical diffusion (Table 1).

\begin{table}[t]
\begin{center}
\begin{tabular}{|c|c|}
 \hline
 
  {\bf \hspace{.5cm}  Ghosts }& {\bf  \hspace{.5cm} Departed Quantities }    \\
 \hline
 
 $\; \frac{\partial U}{\partial t}=D \, \frac{\partial^2 U}{\partial
x^2}$ & \rm{lattice spacing }\\
 & \rm{mean free path/ time/ speed}\\
 
  Wiener Paths &$E[X^{2}_{N}]  \sim O(N)$\\
\hline
\end{tabular}
\caption{ Bishop Berkeley's picturesque description of limits as ghosts has relevance to a statistical mechanical perspective of the diffusion equation.  In the limit (\ref{eqn:D}) the  relevant physical parameters like the mean free path and mean free time `depart' giving us the partial differential equation. The statistical reasoning upon which the equation is based is also a departed quantity and has to be retrieved by counting arguments within statistical mechanics. }
 \label{detab}

\end{center}
\end{table}

While this has been well-known in the statistical mechanics community since Einstein's paper on  Brownian motion in 1905, consider for comparison the equation:
\begin{equation}
\frac{\partial \psi}{\partial t} =i D \frac{\partial^{2}\psi}{\partial x^{2}}.\label{eqn:schrod}
\end{equation}

\noindent  Despite its evident similarity to the diffusion equation,  Schr\"{o}dinger's equation is still, after nearly a century, a remnant of physical or statistical principles that have yet to be identified. We have the ghost so to speak, but the identities of the departed quantities remain elusive to the extent that it is typical to  declare the equation as {\em fundamental!} SGW and I considered this problem in 1982 resulting in a paper on ``Fractal Spacetime''\cite{gord83}. The paper demonstrated that the relation between familiar quantum   equations and their classical counterparts (Table [\ref{detab}]) was deeper than commonly supposed. 
 The geometric picture presented allowed  a statistical perspective on quantum  characteristic lengths  such as the deBroglie and Compton wavelengths and shed some light on the failure of particle conservation on fine scales. It also pointed out that the Heisenberg uncertainty principle, while generally considered a harbinger of quantum mechanics, had a classical analog that is ubiquitous in the presence of random walks. Just as one could see the ghosts of classical statistical mechanics in the quantum equations, one can apparently see the ghosts of quantum mechanics in classical statistical mechanics.

\begin{table}[tb]
\begin{center}
\begin{tabular}{|c|ll|}
 \hline
 & {\bf \hspace{.5cm}  Classical }
  & {\bf  \hspace{.5cm} Quantum }  
     \\
 \hline
 First Order & \(\;  \frac{\partial {\bf U}}{\partial t}= c\, \sig_z
                         \frac{\partial
                       U}{\partial z} + a\, \sig_x U\) & \hspace{-4mm} $\; \,\frac{\partial
                       \Psi}{\partial t}= c\, \sig_z  \frac{ \partial 
                      \Psi }{\partial z}+\Red i \Black \, m\,\sig_x \Psi$    \\
 Second Order & $ \frac{\partial^2 U}{\partial t^2} =
c^2\frac{\partial^2 U}{\partial z^2}
                       +a^2 U\quad$ &$\frac{\partial^2 \psi}{\partial t^2} =
                       c^2\frac{\partial^2
                      \psi}{\partial z^2}  + (\Red i \Black \, m)^2 \psi$   \\
 `Non-relativistic' & $\; \frac{\partial U}{\partial t}=D \, \frac{\partial^2 U}{\partial
x^2}$ & \hspace{1mm}  $\;\; \frac{\partial \psi}{\partial t}= \Red i \Black \,D \,\frac{\partial^2
\psi}{\partial x^2}$  
\\
\hline
\end{tabular}
\end{center}
\caption{ The relation between Classical PDE's based on stochastic models  and their `Quantum' cousins. Formal analytic continuation
 transforms one to the other, but  the stochastic basis for the equations is then missing. Note that the Quantum equations are regarded as {\em fundamental}, the Classical equations are explained by statistical mechanics and the domain of applicability is known. }
 \label{detab}
\end{table}

 While Fractal Spacetime was an exciting excursion into the quantum-statistical-mechanics analogy, it highlighted deeper questions that were, and still are, unanswered within conventional theory\footnote{Such questions go deeper than simply looking at  `interpretations'. They were seriously addressed by the discoverers of quantum mechanics, particularly Schr\"{o}dinger and Dirac. Since that time, a preference for calculation over insight  has dominated the pedagogy of Quantum Mechanics.  There has however been a resurgence of interest in  questions of origin, partly driven by the tantalizing possibility of quantum computing.  `Emergent quantum mechanics' is itself an emerging field that looks for precursors of quantum mechanics.\cite{Emerqm,Emerqm13,EMQM15}.}.
Perhaps the most obvious question is  the origin of the `wavefunction' itself. Regardless of the geometric suggestiveness of FST and the elegance and precision of conventional formulations of quantum mechanics, there remains no answer to the question ``Why does Nature choose linear superposition of square roots of Probability Density Functions rather than the superposition of PDFs themselves?'' There can be no question that Nature {\em does} make this choice in the micro-world. However, until we understand why, quantum mechanics remains an algorithm that we are forced to refine and explore empirically. (For example we can only guess at its limitations in terms of scale and complexity. It clearly applies to point particles in simple environments, it does not apply to cats.)

In this paper we reexamine the relationship between special relativity, classical statistical mechanics and quantum mechanics. The aim is to disentangle them in order to get a clearer picture of a possible origin of the wavefunction. Ultimately the quest is to find the departed quantities that can help explain the ghosts that are the Schr\"{o}dinger and Dirac equations.  The arguments  follow an earlier sketch in reference [7], but are self-contained and more focused on a bridge between classical statistical mechanics and Minkowski space through the introduction of the `history-map'. 

In the first section we have a look at a `smoking gun'; easily accessible evidence that  classical special relativity {\em misses}  quantum mechanics by {\em assuming} a particular continuum limit associated with worldlines. By questioning a double slit experiment in the context of particles that have discretely  marked worldlines, effectively functioning as  discrete clocks,  we find that a combination of special relativity and classical statistical physics is sufficient to discover interference fringes and a precursor of the wavefunction. The precursor, dubbed a `history-map', is a wave-like object associated with a single particle. It relates a particle's actual history to a pattern in space that is not itself a probability density function. Its square is however a PDF associated with the single particle, suggesting the presence of a precursor to Born's postulate and giving a specific analog of wave-particle duality within special relativity. This is all easy to demonstrate visually and its virtue lies in the fact that it is a direct consequence of the relativity postulates with the single addition of marked worldlines. 

In the subsequent section we have a closer look at  `two-bit' clocks in a two dimensional spacetime. We
reexamine the simple arguments that produced the diffusion equation (\ref{eqn:diffn})  and show that an appropriate modification, implementing the history-map induced by special relativity, produces the Schr\"{o}dinger equation.

We conclude with a discussion of the potential benefits of pursuing the statistical mechanics of history-maps.  
 \begin{figure}[t]
   \centering
   \includegraphics[width=0.4\textwidth]{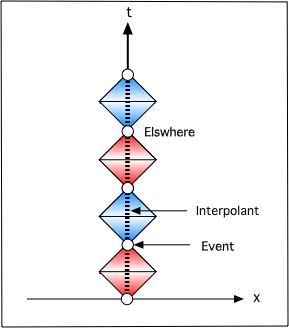} 
     
       \caption{Countable Events, Spacetime Areas and Multiple-Bit Clocks. Here the events (open circles) lie along the worldline in the particle's rest frame. The causal areas between events all have the same Euclidean area and the alternating colour between successive events can be thought of as a binary signal over the worldline of the particle. In this and all subsequent spacetime diagrams, $t$ is vertical and $c=1$.\cite{CC0}}
        \label{fig:count1}
\end{figure}

\section{Special Relativity and  Classical Statistical Mechanics}

We shall be considering physics in one space and one time dimension. Statistical mechanics is all about counting,  so we are going to modify special relativity slightly by avoiding the requirement that `events' be points in spacetime continuously forming the worldline of a particle. Instead, we shall initially assume that there are discrete events along the worldline that appear periodially at the Compton frequency. Between the events are a succession of causal areas that define the time-like regions between events \cfig{fig:count1}. We shall assume that each causal area is naturally partitioned into two areas by its geometry, and that each event distinguishes between local past and future\footnote{In this section we shall argue by images. For simplicity we use two colours to distinguish successive causal regions between events. The two colours only require one `bit' of information to distinguish the two and this will enable us to see half the information we shall ultimately need. The other half of the information, requiring another bit is obtained by distinguishing the two halves of the causal areas.}. 
The succession of events on the worldline then partitions the worldline into four states, however for visual purposes, we shall  use only  two states represented by two colours.

\begin{figure}[t] 
\begin{minipage}{.45\textwidth}
  \centering
  \includegraphics[width=.85\textwidth]{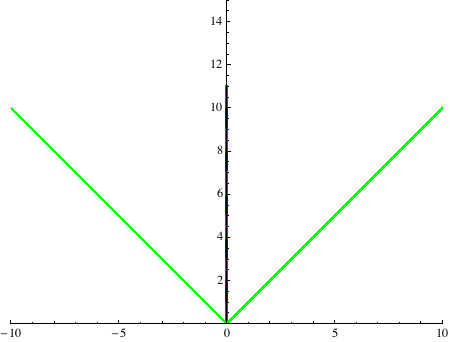}
   \caption{A world-signal, creation at the origin and annihilation at $t=10$. The two colours represent a projection of successive causal areas onto the worldline.\cite{CC0} \textcolor{white}{More text to balance figure sizes.\cite{CC0}}}
   \label{fig:TwoBitPath}
\end{minipage}%
\qquad
\begin{minipage}{.45\textwidth}
  \centering
 \includegraphics[width=.85\textwidth]{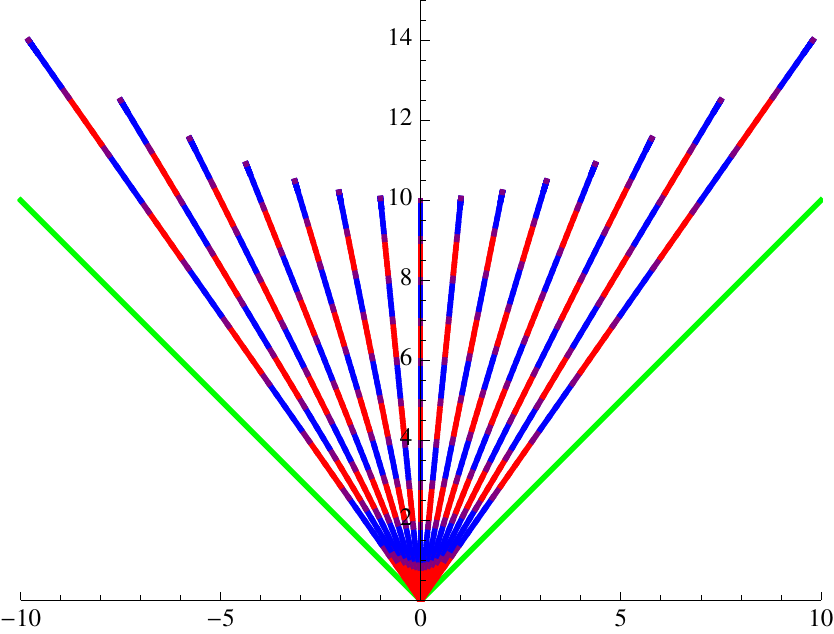}
   \caption{Lorentz boosts of the world-signal in the previous figure. Note that time dilation is `visible' on the time-scale of unity and not just over the time between creation and annihilation.\cite{CC0}}
   \label{fig:ManyTwoBitPaths}
\end{minipage}
\end{figure}

If we colour the causal areas  and consider the projection of the colour onto the worldline in the rest frame we see a `world-signal' associated with the particle in its rest frame \cfig{fig:TwoBitPath}.
If we then consider the images of the world-signal from boosted frames aligned at the origin we see the rest images rotated and stretched by the Lorentz transformation \cfig{fig:ManyTwoBitPaths} that preserves the causal area between events. The binary colouring of the worldsignal here begins to show a characteristic pattern on spacetime. 
The stretching aspect of the transformation is a manifestation of relativistic time dilation. By comparison, the Galilean transformation maintains the independence of space and time and does not stretch the pattern in the same way. The colouring of the one-dimensional subspaces via a Galilean transformation would just translate the colouring from the rest frame horizontally, replacing the hyperbolic pattern of coloured regions with horizontal stripes. 

\begin{figure}[!t] 
\begin{minipage}{.45\textwidth}
  \centering
  \includegraphics[width=\textwidth]{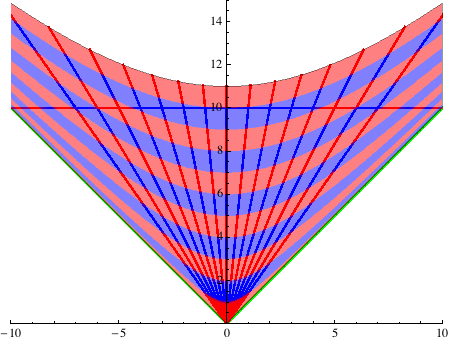}
   \caption{A world-signal with images over many velocities superimposed. The resulting `wave' pattern displays the familiar hyperbolae of fixed proper time. A line parallel to the $x$-axis at $t=10$ samples the one-dimensional subspaces at different points in the particle's history.\cite{CC0}}
   \label{fig:TwoBitProject}
\end{minipage}%
\qquad 
\begin{minipage}{.45\textwidth}
  \centering
  \includegraphics[width=\textwidth]{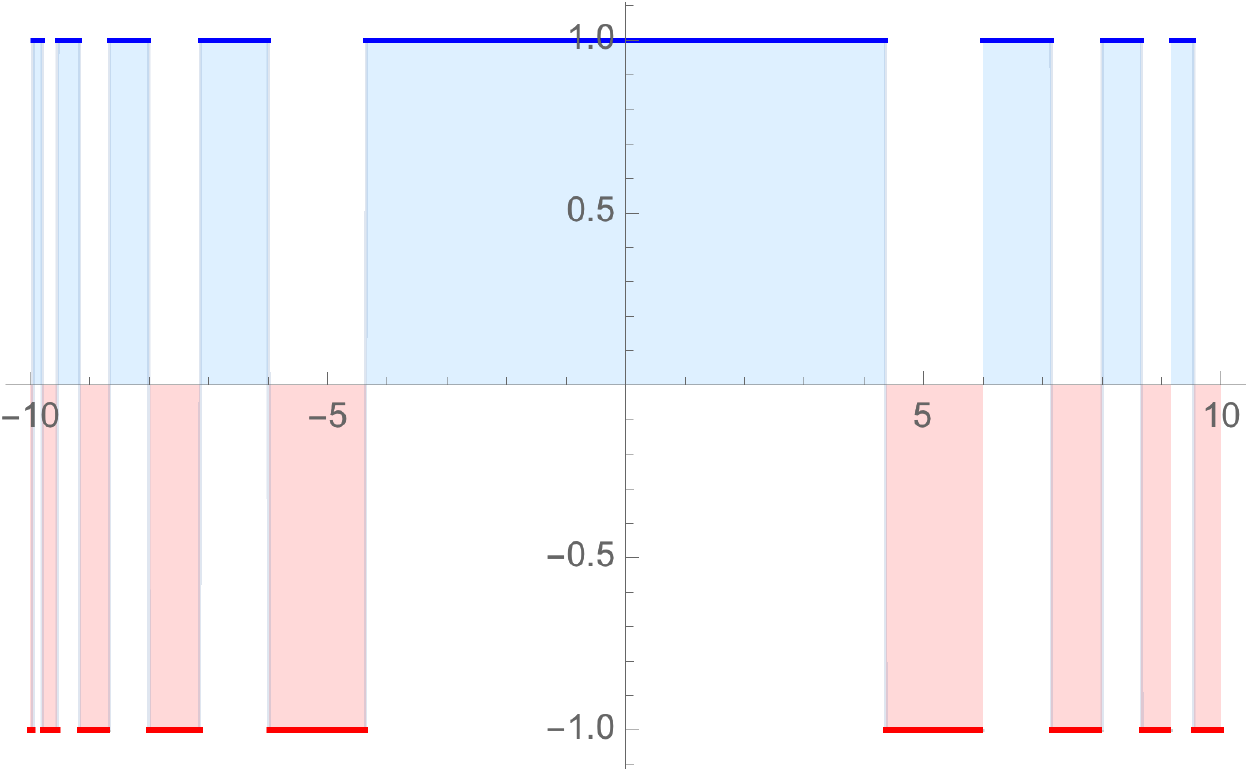}
   \caption{A binary clock at fixed time, $t=10$ in the previous figure.  Here blue is represented by +1,  red by -1. We call such graphs  `History-Maps'. Notice that as you progress out along the $x$-axis from the origin towards the light-cones you encounter the colour states from earlier in the particle's proper time.\cite{CC0} 
   }
   \label{fig:ProjectPaths}
\end{minipage}
\end{figure}

In \cfig{fig:TwoBitProject} we see the `wave-pattern' that emerges if we allow all possible relative velocities. Note that if we sample that pattern at fixed $t$, the colour alternates as a function of $x$. The reason for this is that as you progress out along the $x$-axis towards the light cones, you sample the worldsignal at successively earlier proper times. The resulting function of $x$ we call a `history-map'. An example is sketched in \cfig{fig:ProjectPaths}. Notice that the map itself, while related to the ensemble of inertial frames synchronized at the origin with the rest frame of the clock, {\em is not an ensemble average over many clocks.}  There is a single clock, but a sample of many possible views of the one signal. Each view projects a different point in the clock's actual history onto the $x$-axis. The existence of the pattern is a direct manifestation of time dilation and does not appear in an ensemble of views via the Galilean transformation.

\begin{figure}[b] 

  \centering
 \includegraphics[width=0.5\textwidth]{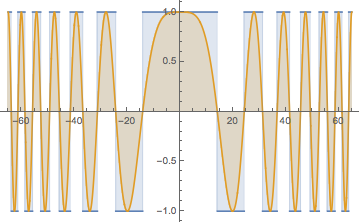}
   \caption{The History-Map of a binary clock is the variable frequency square wave in this figure. The smooth curve is  the real part of the Feynman propagator. A single number...mass, registers the two patterns. The path integral assembles the propagator from an infinite number of Feynman paths. The History-Map is a mapping of a single path history via the Lorentz transformation.}
   \label{fig:BitVsTrig}

\end{figure}

The history-map in \cfig{fig:ProjectPaths} looks a little like a square wave wrapped on a cylinder. The increasing frequency with $x$ is reminiscent of the real part of the Feynman propagator\cite{FeyHibbs}. To test the similarity we look further out in $t$ in the pattern of \cfig{fig:TwoBitProject} so we can have a closer look at the `non-relativistic' regime where the boost velocities are relatively small. The result is shown in \cfig{fig:BitVsTrig}. The history-map is the discontinuous curve in blue and the real part of the Feynman propagator is the continuous curve in light brown. The curves are synchronized on the basis of a single parameter (the particle mass in units of Plank's constant) and it is interesting to note that the synchronization appears exact, at least to several decimal places for small velocities. The history-map seems to `know' where all the nodes and extrema of the Feynman propagator are, despite being a manifestation  of special relativity in a purely classical context.

To be clear that the two curves in \cfig{fig:BitVsTrig} exist on either side of the quantum divide, note that the history-map is a direct consequence of Einstein's postulates with the sole addition of a binary clock attached to worldlines. The history-map in the figure could easily have been drawn back in 1905 with the advent of special relativity. It is also evident  that the square of the signal, as drawn, is a constant function on the interval $x\in [-ct, ct]$ and thus could be considered linked to the uniform PDF for particle detection, presaging the Born postulate. The ghostly appearance of the history-map curve on the wrong side of the classical-quantum boundary suggests we look closely to see if the history-map itself can shed some direct light on  wave-particle duality, a feature of quantum mechanics that is at the heart of its peculiarities.

\begin{figure}[b] 
\begin{minipage}{.45\textwidth}
  \centering
  \includegraphics[scale=0.2]{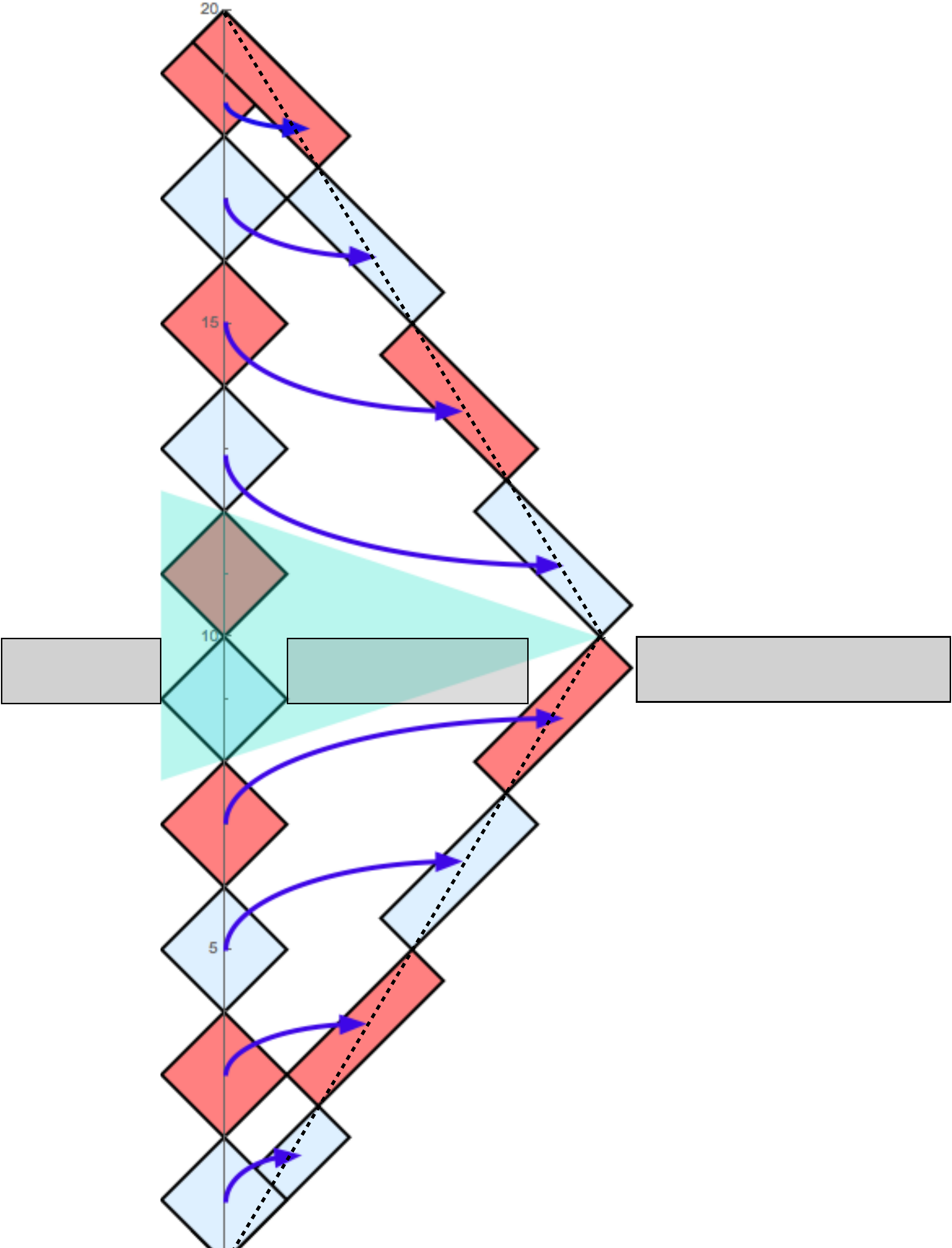} 
   \caption{A clock particle with a barrier that filters paths. The right-hand path is a Lorentz image of the left hand path up to and after the barrier, despite the proper time of the paths being different.\cite{CC0} \textcolor{white}{More text to balance figure sizes. blah blah blah blah blah blah}}%
   \label{fig:HingedSync}
\end{minipage}%
\qquad \qquad
\begin{minipage}{.45\textwidth}
  \centering
 \includegraphics[height=2.2in]{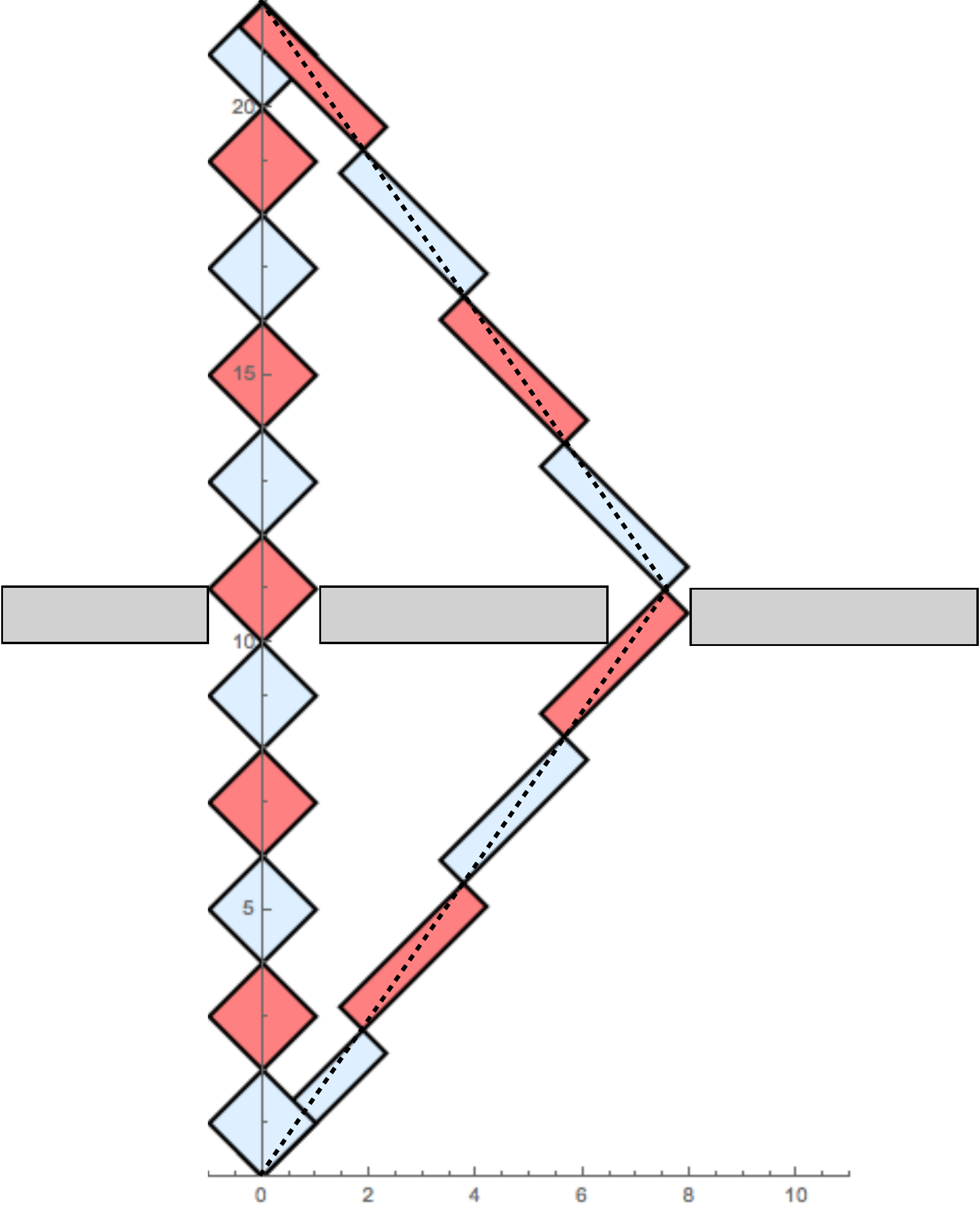} 
   \caption {A clock particle with a barrier that filters paths. The right-hand path is a Lorentz image of the left hand path up to the barrier. After the barrier the right and left paths are not Lorentz images of each other so they have to correspond to separate clocks.\cite{CC0} }
   \label{fig:HingedNoSync}
\end{minipage}
\end{figure}

If the history-map really is a ghost of quantum mechanics then, ideally, we should be able to conjure up a circumstance in which this becomes visible. The canonical context where quantum mechanics comes into conflict with classical probability is in the double slit experiment. In that experiment we have to consider pairs of piecewise inertial frames. In \cfig{fig:HingedSync} we see an inertial frame paired with a hinged frame\footnote{We are assuming that the acceleration required to switch frames may be `hidden' in a time interval small in comparison to the proper time between the ticks of the clock.}, the paths crossing at source and detector screen. In the case depicted, the pair of images begin and end with the same colour. Near the source the two clocks are Lorentz images of each other. Near the detector where they cross they are also Lorentz images of each other. From the sketch we see that the time dilation experienced over the right-hand path is just sufficient to omit a single full period of the stationary clock so that at both ends of the paths the pair are Lorentz equivalent.

In the pair of paths sketched in \cfig{fig:HingedNoSync} the pre-barrier paths are Lorentz images from the source but at the screen the colour is out of phase so the left and right paths are not images of the same clock-particle. The relativity postulate requiring the equivalence of inertial frames does not give us specific direction on piecewise inertial frames. However note that  the patterns we detect in figures 2 through 6 all arise from {\em images} of a single clock-particle's history over an {\em ensemble} of inertial frames specified by the relativity postulate. The pair of paths in \cfig{fig:HingedSync} are boost images of each other based on the Lorentz transformations from either end, whereas the paths in  \cfig{fig:HingedNoSync} are clearly not images of the same clock-particle. This strongly suggests we should consider the paths in  \cfig{fig:HingedSync} as equivalent whereas the paths in  \cfig{fig:HingedNoSync} would be inequivalent and should be eliminated as pairs of paths from ensembles of equivalent frames. 

On this basis let us consider the analog of the Young double slit experiment where we allow only pairs of paths that satisfy the above Lorentz equivalence. 

\begin{figure}[!tbp] 
    \centering
     \includegraphics[scale=.4]{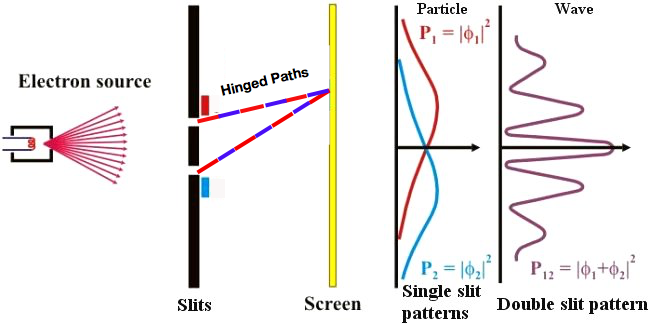} 
    \caption{A schematic of the Young double slit experiment for clock-particles.\cite{CC0}}
    \label{fig:young}
 \end{figure}

  \cfig{fig:young} is a sketch of the double slit experiment. Here we imagine a source of individual non-interacting  clock-particles. Imagine the  paths shown emanating from the emitter to   be just Lorentz images of a {\em single actual electron path}. The two slits filter the boost images of the `actual' path and at each point on the screen we imagine the two incoming images from the two slits. If we ignore the Lorentz equivalence  a probabilistic argument would work and we would expect, over many experiments,  a `particle' pattern to emerge.
  
  However, suppose at the screen we require the incoming paths to be Lorentz equivalent in order for the actual particle to land there. This would be forcing the `equivalence of inertial frames' to make sure that the pair of paths were

  \noindent Lorentz images of each other. We might expect regions of the screen to be inaccessible to our clock particles as a result.

 We can test this by using the history-map of a particle that oscillates between $\pm 1$ to implement the exclusion of inequivalent paths.  
\cfig{fig:XORpic} illustrates the filtering condition in terms of colour. 
 Table[\ref{tab:booktabs}] illustrates the use of averages of Ising spin variables to implement the filter.

 Using the same mass parameter for the history-map to register the image in \cfig{fig:BitVsTrig}, we can add the two history-maps for the double slit and compare the result with that obtained from the Feynman propagator.

 \vspace{\baselineskip}
 {\centering \fbox{
\begin{minipage}{.45\textwidth}
  \centering
 \fbox{\includegraphics[scale=0.3]{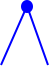}} \quad Yes\\

\fbox{\includegraphics[scale=0.3]{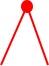}} \quad Yes\\

\fbox{\includegraphics[scale=0.3]{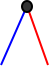}} \quad No\\

   \captionof{figure}{The use of $\sigma=\pm1$ in counting paths to implement an exclusive OR in parity between pairs of paths. In terms of the colour used in figures [7-9] this provides a filter for Lorentz equivalent paths.\cite{CC0}}
   \label{fig:XORpic}
\end{minipage}%
\qquad \qquad
\begin{minipage}{.45\textwidth}
  \centering

      \begin{tabular}{@{} lcr @{}} 
      \toprule
      \multicolumn{3}{c}{$\chi=(\sigma_{A}+\sigma_{B})/2$} \\
      \cmidrule(lr){1-3} 
      $\sigma_{A}$\textbackslash $\sigma_{B}$   & 1 & -1 \\
      \midrule
      \; ~1     & 1 & 0 \\
     \;  -1      & 0  & -1 \\
       
      \bottomrule
   \end{tabular}
  \captionof{table}{The XOR for pairs of paths using the `Ising spin' variable $\sigma=\pm1$. Notice that $\chi^{
   {2} }\in \{ 0,1\}$  is the usual binary characteristic function used for counting in probability!}
   \label{tab:booktabs}
\end{minipage}
}
}

  \begin{figure}[h!] 
   \centering
    \includegraphics[scale=0.33]{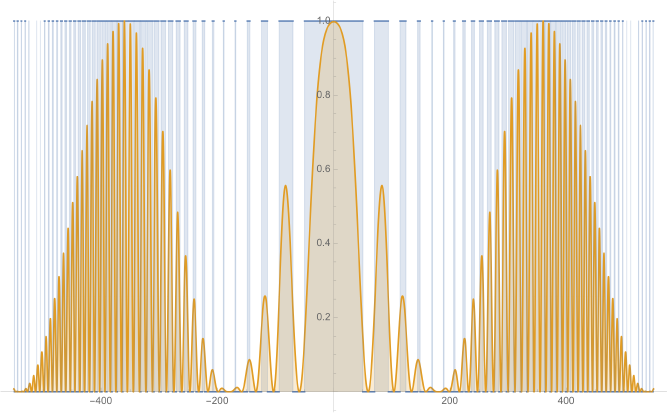} 
   \caption{The contribution to the probability of detection from the real part of the Feynman propagator to the three central fringes of an idealized double slit experiment. The light blue area is the same experiment using the History-Map generated by the parity filter. The propagator and History-Map both pick out the interference fringes. The Feynman calculation is quantum mechanics. The History-Map calculation is special relativity with worldlines marked to carry digital world-signals.}
   \label{fig:fringe}
\end{figure}
   The result is shown in \cfig{fig:fringe} for the central three fringes. As in the direct comparison of Feynman's propagator and the history-map  \cfig{fig:BitVsTrig}, it appears that the addition of two history-maps produces a digital version of the propagator result, since the nodes appear to be precisely registered. 
This result confirms at a qualitative level that attaching a discrete clock to worldlines and extending the relativity principle to the discrete signal produces a result that closely imitates the `wave-particle' duality of quantum mechanics.

 It is  again important to note that \cfig{fig:fringe} compares results that are either side of the quantum-classical divide. The smooth curve is routine quantum mechanics, the discontinuous curve is classical special relativity with single-bit discrimination between events at the Compton wavelength.

Perhaps the most interesting observation from this is the foreshadowing of a superposition principle for wavefunctions. The history-map interacts in the above context via the addition in Table[3]. Pairs of paths add in a binary fashion in that they either agree and are counted or disagree and are annihilated from the ensemble of equivalent signals. In the latter case, removal from a probabilistic ensemble provides just the sort of pre-processing of probability that is provided by wavefunctions in the quantum context. The difference here is that we see {\em explicitly} the origin and function of  the pre-processing. Simply put, the superposition of binary worldsignals implements the relativity postulate generalized to allow for binary signals at the Compton scale. This compares favourably with the superposition of wavefunctions in the quantum case, where the principle  follows from the linearity of the appropriate differential equations, but is difficult to associate directly with a method of counting\footnote{Here again Berkeley's  reference to ghosts haunts the discussion of partial differential equations. Einstein had to step outside the diffusion equation and invoke ideas from statistical mechanics to demonstrate that the diffusion equation was consistent with, and given plausible assumptions, actually followed from the atomic hypothesis. Einstein had to find the 'departed quantities' that existed prior to a continuum limit and had to decide what, exactly, to count. Here we are beginning to see a similar situation for the Schr\"{o}dinger equation. We need to find what the departed quantities are and how to arrange counting {\em before} taking a continuum limit.}.

\section{Averages of History-Maps and Schr\"{o}dinger's Equation}

In the previous section we uncovered some ghosts of quantum mechanics haunting special relativity. The ghosts themselves are vanquished in conventional relativity by the worldline assumption that replaces a digital worldsignal by a constant function. This removes mass from the kinematics and renders time dilation {\em invisible}  and irrelevant on small scales.

\begin{figure}[tbp] 
   \centering
   \includegraphics[scale=0.25]{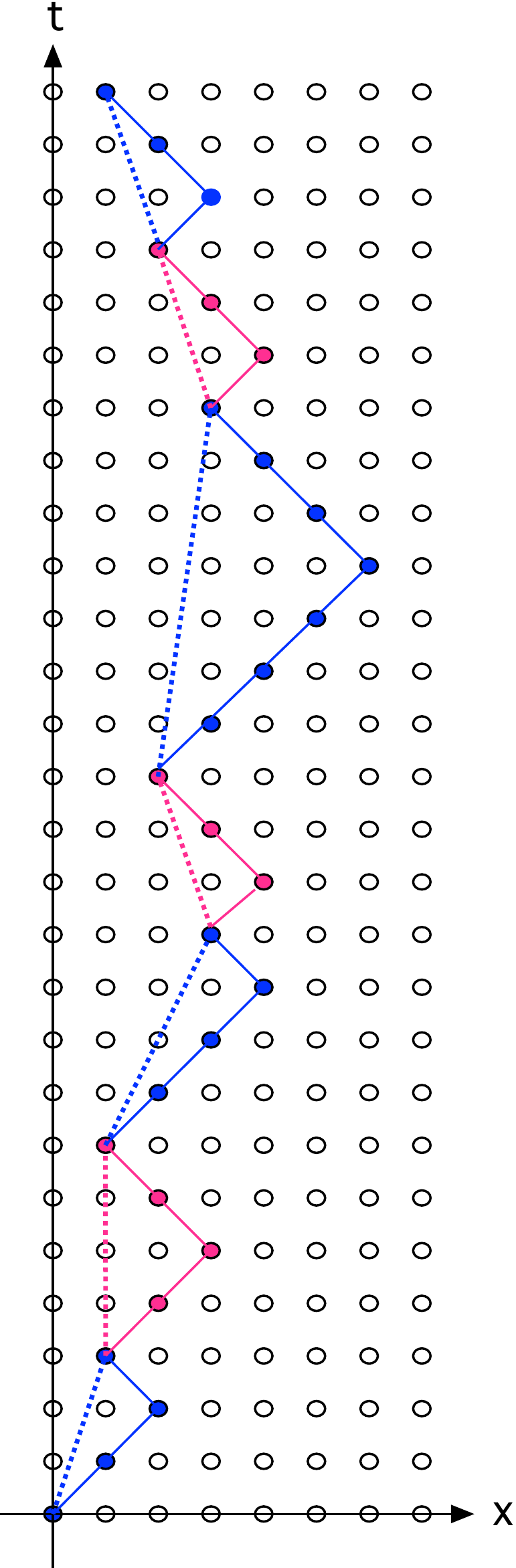} 
   \caption{A path on a square lattice. The path has binary colouring that changes with every two corners, mimicking the colour change at events on the worldline illustrated in Fig.[2]. The approximation to the worldline is the piecewise linear path between every other corner, corresponding to the dashed lines in the figure. The `light-cone' paths serve to colour the analog of the world-line and essentially partition it into four possible states.\cite{CC0}}
   \label{fig:EnvelopePath2}
\end{figure}

If we are to uncover the Schr\"{o}dinger equation from special relativity in a manner similar to our simple route to the diffusion equation [1] through  [2-6], we need to revisit the way paths  contributed to the diffusion equation and modify the path counting based on history-maps. Consider \cfig{fig:EnvelopePath2}. 
The sample path illustrated would, in the diffusive case, have a single positive weight proportional to $(1/2)^{n}$ at its end-point, where $n$ is the number of steps in the path. All such paths would have the same positive weight and the difference equation (\ref{eqn:mass}) requiring (probabilistic) mass conservation determines this.

The history-map, in contrast, does not conserve mass. The history-map partitions paths into four states based on causal areas and uses the four states to {\em filter out paths that are not Lorentz equivalent}. If we are to regard the path in \cfig{fig:EnvelopePath2} in terms of a history-map, we notice that the colour of the path changes for every two corners in the path, and that there are two  possible path directions at each step\footnote{This is an example of Feynman's `corner rule' \cite{GNOIJTP2010} that is a feature of his Chessboard (Checkerboard) model.\cite{FeyHibbs} \cite{Kauffman:1996kq}\cite{OrdAnnals09}}. We shall have to keep track, not only of the path but of the `state' of the path at each step. In order to do this we shall have to record more of the history of paths than in the case of the diffusive model.

\begin{figure}[!t] 
\begin{minipage}{.55\textwidth}
  \centering
   \includegraphics[width=\textwidth]{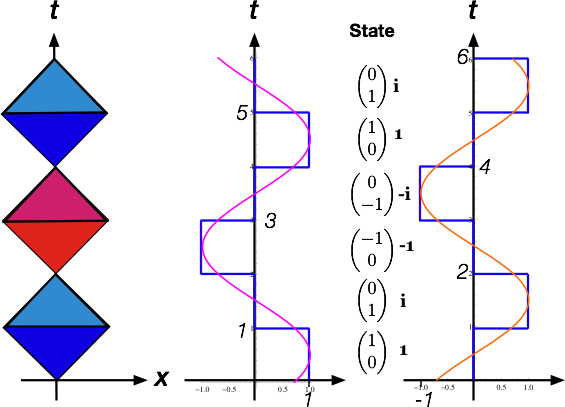} 
   \caption{The Two-Bit clock states. Two successive causal areas naturally partition into four areas. The boundaries of the areas themselves are parallel to the two light-cones that form an invariant coordinate system for the clocks. The analog of the Ising spin variable that discriminates between even and odd is a two-component state, listed between the two component signals in the figure. Superimposed on the component signals are what might be expected as a result of including a stochastic feature to arrive at wavefunctions.\cite{CC0}}
\label{fig:EncodePath}
\end{minipage}%
\qquad
\begin{minipage}{.35\textwidth}
  \centering
  \includegraphics[width=\textwidth]{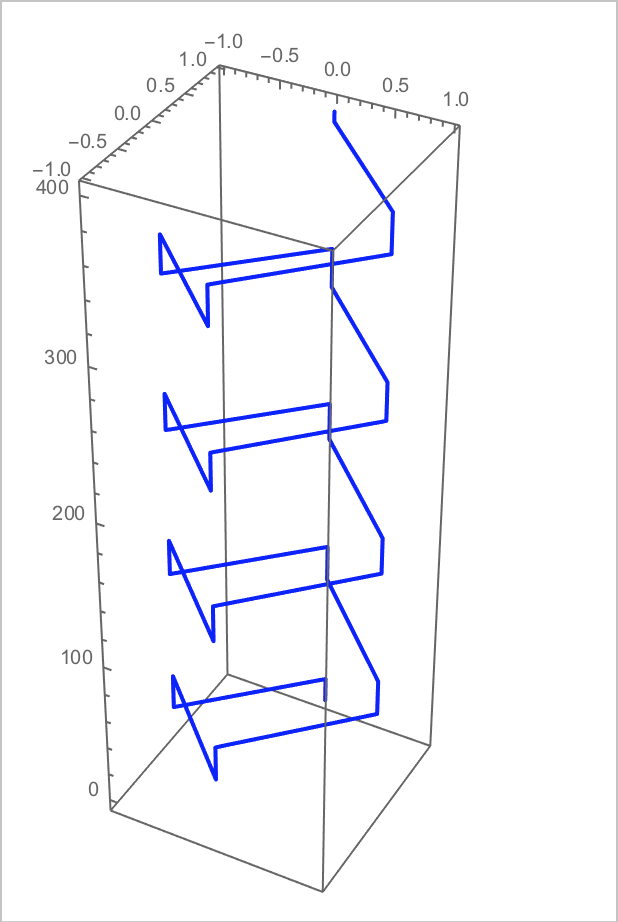} 
   \caption{The two-bit clock from the figure to the left as a square spiral. The discrete two-bit nature of the spiral allows the discrete signal to create a history-map to serve as a filter, eliminating paths that are not Lorentz images of each other.\cite{CC0}}
   \label{fig:SquareSpiral}
\end{minipage}%
\end{figure}

To pursue a quantitative version of the previous section we need to set up some machinery for counting paths. \cfig{fig:EncodePath} and \cfig{fig:SquareSpiral} illustrate an option for doing this.
Our two-bit  clocks requires four states that we can specify in a set $S_{C}$ of two component objects. Each component is either absent, or takes on a value $\pm 1$. 
\begin{equation}
S_{C}=\{\left(\begin{array}{c}1 \\0\end{array}\right),\left(\begin{array}{c}0 \\1\end{array}\right), \left(\begin{array}{c}-1 \\0\end{array}\right), \left(\begin{array}{c}0 \\-1\end{array}\right)\}=\{s_{1}, \ldots s_{4}\}
\label{eqn:states}
\end{equation}
Periodic evolution through the four states give a time evolution corresponding to the rectangular spiral  of \cfig{fig:SquareSpiral}.
 
The  spiral of \cfig{fig:SquareSpiral} is the two-bit analog of the one-bit signal that we represented by two colours in \cfig{fig:count1}, using the two colours to produce the history-map \cfig{fig:ProjectPaths}. In \cfig{fig:BitVsTrig} we saw that the Feynman Propagator looked visually like a softened version of the history-map that we might suspect would emerge from a statistical averaging. We might anticipate that in the two-bit case  \cfig{fig:SquareSpiral}, the statistical averaging might serve to soften the square spiral locally into a circular spiral. This is indeed what happens as the following calculations show.

The four states of the discrete spiral (\ref{eqn:states})
 satisfy cyclically \begin{equation}
 s_{k+1}=\sigma_{C}s_{k}
 \end{equation}
  where \[\sigma_{C}=\left(\begin{array}{cc}0 & -1 \\1 & 0\end{array}\right).\] The states $s_{1}$ and $s_{3}$ are inequivalent as are  $s_{2}$ and $s_{4}$.

The digital clock of \cfig{fig:EncodePath} illustrates how random walk paths like those illustrated in \cfig{fig:EnvelopePath2} might be filtered to eliminate Lorentz inequivalent paths.   To see how this occurs with our diffusive sample paths we track random walks on a square lattice, giving them state changes on change of direction. To this end, define densities of Clock-Particles on a lattice by
 $p_\mu(m \delta, s \epsilon)$, 
$(\mu = 1,2,3,4)$. These  represent the probabilities
that a C-particle leaves a
space time point $(m \delta, s \epsilon)$  in state $\mu$
$(m= 0,\pm 1,\ldots;\;s=0,1,\dots)$.
The difference equations for $p_\mu$  are

\begin{equation}
\begin{array}{rcl}
p_1(m\delta,(s+1)\epsilon) & = & \frac{1}{2} p_1 ((m-1)\delta,s \epsilon)
 + \frac{1}{2}
p_4((m+1)\delta, s \epsilon) \\
p_2(m\delta,(s+1)\epsilon) & = & \frac{1}{2} p_2 ((m+1)\delta,s
\epsilon) + \frac{1}{2} p_1((m-
1)\delta, s \epsilon) \\
p_3(m\delta,(s+1)\epsilon) & = & \frac{1}{2} p_3 ((m-1)\delta,s
\epsilon) + \frac{1}{2}
p_2((m+1)\delta, s \epsilon) \\
p_4(m\delta,(s+1)\epsilon) & = & \frac{1}{2} p_4 ((m+1)\delta,s
\epsilon) + \frac{1}{2} p_3((m-
1)\delta, s \epsilon).
\end{array}
\end{equation}

\noindent These equations express the conservation of the number of particles over time, with half of them maintaining their direction and state at a time step and half changing direction and state. This is simply a generalization of eqn.(2) that can allow us to keep track of state changes as well as position. The four states refer to the two spatial directions and the two possible `states' for each step.

We can express the coupled equations in matrix form with shift operators
$E_x^{\pm1}$ and $E_t$ such that
\begin{eqnarray}\nonumber
E_x^{\pm1} \; p_i(m \delta, s \epsilon) &=& p_i((m \pm 1)\delta, s
\epsilon)\quad  {\rm and}\\ \nonumber
E_t \quad p_i(m \delta, s \epsilon) &=&p_i(m\delta, (s+1)
\epsilon).\end{eqnarray}
The difference equations may then be written as
$$E_t \; P(m \delta, s \epsilon) = \frac{1}{2}
\left[\begin{array}{cccc}E_x^{-1}&0&0&E_x\\
E_x^{-1}&E_x&0&0\\
0&E_x&E_x^{-1}&0\\
0&0&E_x^{-1}&E_x \end{array}\right] P(m
\delta, s\epsilon) $$
\addtocounter{equation}{1}
\noindent where
$P(m \delta, s \epsilon) $ is a column vector of the $p_\mu$.

Now consider a change of variables:

\begin{equation}
z_1 =\frac{p_1 +p_3}{2},\;z_2 =\frac{p_2+p_4}{2}\label{eqn:cofz}
\end{equation}

\noindent and
\begin{equation}
 \phi_1 =\frac{p_1 -p_3}{2},\;\phi_2 =\frac{p_2 -p_4}{2}.\label{eqn:cofv}
\end{equation}

\noindent The $z_{k}$  just represent probabilities, partitioned by direction.  The $\phi_{k}$ record the net number of paths that are `Lorentz equivalent' using the $\pm 1$ filtering process   illustrated in \cfig{fig:XORpic} and Table 1.

 The change of variables block diagonalizes the shift matrix to give:
\begin{equation}
E_t \left[\begin{array}{c}z_1\\ z_2\\ \phi_1\\
\phi_2 \end{array}\right] =
\frac{1}{2} \left[\begin{array}{cccc}E_x^{-1}  &E_x
 &0&0\\
E_x^{-1}&E_x&0&0\\
0&0&E_x^{-1}&-E_x\\
0&0&E_x^{-
1}&E_x\end{array}\right]\;\left[\begin{array}{c}z_1\\ z_2\\
\phi_1\\ \phi_2
\end{array}\right].\end{equation}
The upper block gives a discrete form of the diffusion equation.

\begin{equation}
E_t \left[\begin{array}{c} z_1\\
z_2 \end{array}\right] =
\frac{1}{2} \left[\begin{array}{cc}E_x^{-1}&E_x\\
E_x^{-
1}&E_x\end{array}\right]\;\left[\begin{array}{c}
z_1\\ z_2
\end{array}\right] 
\label{eqn:shift}\end{equation}  the lower block is:
\begin{equation}
E_t \left[\begin{array}{c} \phi_1\\
\phi_2 \end{array}\right] =
\frac{\alpha}{2} \left[\begin{array}{cc}E_x^{-1}&-E_x\\
E_x^{-
1}&E_x\end{array}\right]\;\left[\begin{array}{c}
\phi_1\\ \phi_2
\end{array}\right] 
\label{eqn:shift0}\end{equation} 
where  a normalization constant $\alpha$ has been inserted. This is required since this is a filtering process that will eliminate many paths.  However the ensemble of filtered paths `live' in different eigenspaces so we can choose to track only the filtered paths.

If we try to apply the limit (\ref{eqn:D}) directly to (\ref{eqn:shift0}), the result does not converge to give us a differential equation. The difference equations (\ref{eqn:shift0}) change state with every direction change. In the limit (\ref{eqn:D}) the frequency of those direction changes goes to infinity, corresponding to an unbounded increase in the `zitterbewegung' arising from the rest mass\footnote{Comparing the path in Fig.\ref{fig:EnvelopePath2} with Fig.\ref{fig:count1} we see that the path being tracked outlines the causal areas and in the limit (\ref{eqn:D}) will increase the Compton frequency without bound, sending the rest-mass to infinity.}. However, because the two lattice directions are equally likely, on average there  will be a state change every two steps and a complete return to the original state in eight steps. This suggests that in order to avoid the dominance of the zitterbewegung signal, we look for a Taylor expansion of eqn. (\ref{eqn:shift0}) over eight time steps rather than just one. 
Thus we consider expanding

\begin{equation}
(E_t^{8}-1) \left[\begin{array}{c} \phi_1\\
\phi_2 \end{array}\right] =
\left ((\frac{\alpha}{2})^{8} \left[\begin{array}{cc}E_x^{-1}&-E_x\\
E_x^{-
1}&E_x\end{array}\right]^{8}-\left[\begin{array}{cc}1 & 0 \\0 & 1\end{array}\right]\right)\;\left[\begin{array}{c}
\phi_1\\ \phi_2
\end{array}\right]. 
\label{eqn:shift1}\end{equation} 

\noindent For $\alpha=\sqrt{2}$  and writing $m \delta =x$ and $s \epsilon =t$, we find that to lowest order we get:
\begin{equation}
\frac{\partial}{\partial t} \left[\begin{array}{c} \phi_1(x,t)\\
 \phi_2(x,t)\end{array}\right] (8 \epsilon)
=
\left(
\begin{array}{cc}
 0 &  1 \\
 -  1& 0 \\
\end{array}
\right)\frac{\partial^{2}}{\partial x^{2}} \left[\begin{array}{c} \phi_1(x,t)\\
 \phi_2(x,t)\end{array}\right] (4 \delta^{2})\label{eqn:s2}\end{equation}
This gives us the coupled differential equations
\begin{equation}
\frac{\partial}{\partial t} \left[\begin{array}{c} \phi_1(z,t)\\
 \phi_2(z,t)\end{array}\right] 
=
D \left(
\begin{array}{cc}
 0 &  1 \\
 -  1& 0 \\
\end{array}
\right)\frac{\partial^{2}}{\partial x^{2}} \left[\begin{array}{c} \phi_1(x,t)\\
 \phi_2(x,t)\end{array}\right]\label{eqn:s2}\end{equation}
 
\noindent  provided we refine the lattice in such a way that $\frac{\delta^2}{2 \epsilon}\to D$. Notice the presence of $-\sigma_{C}$ here. The change of variables (\ref{eqn:cofv}) to the $\phi$ that count the difference in occupation between states has uncovered the operator that switches signs every two corners and advances the two-bit clock \cfig{fig:EncodePath}. 
 
To find  a more familiar form of this equation, write
\begin{eqnarray}\label{eqn:iin}
\psi_+(x,t)&=&\left (\; i\ \phi_1(x,t) + \phi_2(x,t)\, \right)/2\\\nonumber
\psi_-(x,t)&=&(\, -i \ \phi_1(x,t) + \phi_2(x,t)\,)/2,\\\nonumber
\end{eqnarray}
In terms of the $\psi$ eqn.(\ref{eqn:s2}) becomes: 
\begin{equation}
\frac{\partial}{\partial t} \left[\begin{array}{c} \psi_+(z,t)\\
 \psi_-(z,t)\end{array}\right] 
=
D \left(
\begin{array}{cc}
 i &  0 \\
 0& -i\\
\end{array}
\right)\frac{\partial^{2}}{\partial x^{2}} \left[\begin{array}{c} \psi_+(x,t)\\
 \psi_-(x,t)\end{array}\right]\label{eqn:withi}\end{equation}
a pair of conjugate Schr\"{o}dinger equations.  

Notice here that in the change of variables (\ref{eqn:iin}) used to obtain (\ref{eqn:withi}), the use of the unit imaginary $i$ is not an analytic continuation. The path counting in the model naturally produces the (Real) $\phi$, however because  the `switch' $\sigma_{C}$  has eigenvalues $\pm i$, complex amplitudes simplify the representation. As one might expect, the counting arguments we have used to arrive at Schr\"{o}dinger's equation cannot reach outside rational numbers to produce the unit imaginary. However, with the use of the simple filter we have built, they can produce Real functions (ie. the $\phi$) that are attuned to the switch $\sigma_{C}$ and thus ultimately reflect the eigenvalues of that matrix. 
Thus it is worthwhile to  note that despite (\ref{eqn:withi}) there is no formal quantum mechanics here. Equation(\ref{eqn:withi}) just represents a form of `relativistic filtering'. It is a stochastic model of the subtraction of inequivalent paths discussed in association with \cfig{fig:XORpic}.  This is an implementation of a statistical averaging over History-Maps to obtain a continuum limit using a `non-relativistic' approximation.\footnote{ The stroboscopic limit that uses eight steps, eqn.(\ref{eqn:shift1}) eliminates the rest mass component at the Compton frequency and the scaling $\frac{\delta^2}{2 \epsilon}\to D$ sends the mean free speed to infinity essentially leaving the non-relative kinetic energy term intact.}\hspace{-1mm}  
 
Tracking through the above calculation we have found  the Schr\"{o}dinger equation, but more interestingly we can identify several departed quantities in the process. 

\begin{itemize}

\item Perhaps the most important and surprising departure is special relativity itself. It is practically universal in the pedagogy surrounding quantum mechanics that Schr\"{o}dinger's equation is considered `non-relativistic'. The association of the equation with Hamiltonian mechanics reinforces the impression that, for all practical purposes,  the equation itself is independent of relativity. However, in the above view the whole path-counting algorithm is a remnant of special relativity and {\em would not exist without time dilation.} It is after all  time dilation that gives rise to the history-map. There is no analog of this in the Galilean transformation and it is for this reason that canonical quantization in the Schr\"{o}dinger regime is forced to invoke an analytic continuation to bring in phase, the natural phase dependence on velocity in special relativity having been lost in  the transition to Newtonian physics. 

\item Another surprising departure is the relocation of wave-particle duality firmly into the domain of Minkowski spacetime. The astonishing thing about the double-slit experiment is that a particle, observed after having passed through the double slit, has to  have `known' that both slits were open in order to avoid being detected in dark areas of the interference pattern. This seems to violate the whole idea of a particle as a localized entity. We conventionally invoke wave-particle duality to `explain' this. The above formulation does not fully explain the paradoxical association of a wave and a particle but does point {\em directly} to its origin in special relativity. 

The equivalence of inertial frames with $c$ fixed is the antecedent of the difference equation (\ref{eqn:shift0}), just as mass conservation is the origin of the diffusion equation. We do not recognize the presence of wave-particle duality as an intrinsic feature of special relativity simply because we eliminate its local appearance by assuming the worldline is characterized by a constant (scale-invariant) signal. On one hand, relativity acknowledges the central importance of spacetime (``Spacetime tells matter how to move''), on the other hand, invoking a scale invariant worldline {\em removes} signal-processing in general from the theatre of operations (in particular removing the Fourier uncertainty principle) resulting in a theory (special relativity)  that misses quantum mechanics entirely. Reinstating a periodic signal on the worldline implicates signal processing and associates a history-map with each particle. If we believe in the equivalence of inertial frames as formulated in special relativity, the history map as a precursor to a wavefunction is a short step away. The non-locality of quantum propagation in this view is a direct consequence of the relativistic implementation of the equivalence of inertial frames.

\item A third encouraging feature of the history-map view is that the algorithmic nature of wavefunctions is explicit. In conventional quantum mechanics, the wavefunction is part of a calculational scheme to extract eigenvalues and eigenfunctions that appear to have direct relevance in the observation of Nature. The mathematical structure of quantum mechanics has evolved and been refined over the years primarily to ensure mathematical consistency while retaining empirical accuracy.  Our ghosts are better dressed and more refined than when they first appeared, but the denial of the very existence of departed quantities means that, conventionally, we have no insight into what the function of wave amplitudes really is. This picture of their origin strongly suggests that their {\em function} is to act as a relativistic filter to enforce a form of Lorentz covariance on fine scales, even in the `non-relativistic' limit.
\end{itemize}

\section{Discussion}

Berkeley's criticism that the pioneers of the calculus believed in `ghosts' that had no precise mathematical foundation, spurred mathematicians  to change the foundations. Within the statistical mechanics community, there is an increased awareness that the new foundations  are about the ghosts of {\em departing} quantities, with an acknowledgement that the departure itself is of great importance.

The issue is perhaps less appreciated in the physics community at large. When it comes to concepts like space and time, we have no hope of ever physically measuring the full spectrum of $\delta$'s and $\epsilon$'s we use in limiting processes. Yet we routinely accept the calculus as the starting point for fundamental physics. A case in point is 
the model of a particle history as a smooth curve;  a worldline in a spacetime continuum. While intuitive and convenient, the model as specified is indeed a ghost. It inherits the full machinery of modern analysis from the start. However, its incompatibility with the uncertainty principle should give us pause. If we assume that particles have smooth worldlines that are featureless, regardless of mass, then we appear to be confined by classical  relativity and have to accept  quantum mechanics as an overlay on a classical theory. The price we pay for this is the current plethora of `interpretations' of the theory. We accept precision over insight.

The above sketch suggests that there may be another way to proceed. If we replace the worldline with a sequence of worldpoints as per \cfig{fig:count1}, this ultimately allows us to consider a history-map of a particle, a spatial representation of that particle's history. This provides a direct analog of wave-particle duality {\em within special relativity}. At this point in time, it can be shown that this provides two central ghosts, the Schr\"{o}dinger and Dirac free particle equations in  two dimensional spacetimes, complete with departing quantities.  The primary virtue of this is that the `departing quantities' show us that the function of probability amplitudes is to pre-process the ensemble of paths to count only those that are in accord with special relativity.  At a minimum, the resulting superposition of wavefunctions rather than PDFs  and the implication of unitary propagation,  suggest that the formal similarity between the classical and quantum equations in Table[\ref{detab}] have a direct origin within classical statistical mechanics, provided we start in a discrete relativistic framework.

 This work uncovers  a small but representative fragment of quantum mechanics {\em as a recognizable algorithm}. The Dirac equation in two dimensions is easily obtained from this perspective since one can view Feynman's Chessboard model\cite{{Kauffman:1996kq}} as an implementation of this picture. Formal extension of the Dirac equation to four dimensions can be accomplished simply by switching  to the appropriate Clifford algebra from the Chessboard model. 

As this festschrift for SGW is hopefully part of a long sequence of them, it is perhaps appropriate to give Bishop Berkeley the last word.

\begin{aquote}{From Siris: A Chain of Philosophical Reflexions and Inquiries\cite{Berkeley1747}  }{ The eye by long use comes to see even in the darkest cavern: and there is no subject so obscure but we may discern some glimpse of truth by long poring on it.}
\end{aquote}

\section{References}
\bibliographystyle{unsrt}

\end{document}